\documentclass[prb,twocolumn,showpacs,preprintnumbers,amsmath,amssymb,superscriptaddress]{revtex4}

\usepackage{graphicx}
\usepackage{dcolumn}
\usepackage{bm}

\begin{document}

\bibliographystyle{apsrev}

\title{Boundary Slip as a Result of a Prewetting Transition}

\author{Denis Andrienko}

\email{denis@iop.kiev.ua}

\affiliation{Max-Planck-Institut f\"{u}r Polymerforschung, %
Ackermannweg 10, D-55128 Mainz, Germany}

\author{Burkhard D\"unweg}

\email{duenweg@mpip-mainz.mpg.de}

\affiliation{Max-Planck-Institut f\"{u}r Polymerforschung, %
Ackermannweg 10, D-55128 Mainz, Germany}

\author{Olga I. Vinogradova}

\email{vinograd@mpip-mainz.mpg.de}

\affiliation{Max-Planck-Institut f\"{u}r Polymerforschung, %
Ackermannweg 10, D-55128 Mainz, Germany}

\affiliation{Laboratory of Physical Chemistry of Modified Surfaces, %
Institute of Physical Chemistry, Russian Academy of Sciences, %
31 Leninsky Prospect, 119991 Moscow, Russia}

\date{\today}

\begin{abstract}
  
  Some fluids exhibit anomalously low friction when flowing against a
  certain solid wall. To recover the viscosity of a bulk fluid, slip
  at the wall is usually postulated. On a macroscopic level, a large
  slip length can be explained as a formation of a film of gas or
  phase-separated `lubricant' with lower viscosity between the fluid
  and the solid wall. Here we justify such an assumption in terms of a
  prewetting transition. In our model the thin-thick film transition
  together with the viscosity contrast gives rise to a large boundary
  slip. The calculated value of the slip length has a jump at the
  prewetting transition temperature which depends on the strength of
  the fluid-surface interaction (contact angle). Furthermore, the
  temperature dependence of the slip length is non-monotonous.

\end{abstract}


\maketitle

\section{Introduction}


It is accepted in hydrodynamics that the velocity of a liquid
immediately adjacent to a solid is equal to that of the
solid~\cite{batchelor.gk:1967.a}. Such an absence of a jump in the
velocity of a simple liquid at a surface seems to be a confirmed fact
in {\em macroscopic} experiments. However, it is difficult to obtain
the same conclusion using {\em microscopic} models. It has been
noticed that, even in case of simple liquids, the no-slip boundary
condition is not justified on a microscopic level.

\begin{figure}
\begin{center}
\includegraphics[width=5cm, angle=0]{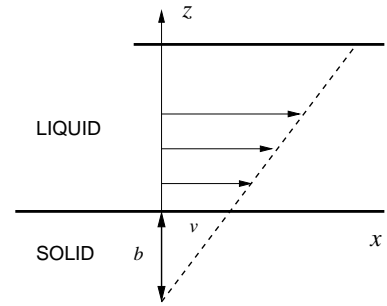}
\end{center}

\caption[slip length]{
\label{fig:slip} 
Definition of the slip length $b$ for a simple shear flow.
}
\end{figure}

Therefore, the conventional condition of continuity of the velocity,
or the {\em no-slip} boundary condition, is not an exact law but a
statement of what may be expected to happen in normal
circumstances. While the normal component of the liquid velocity must
vanish at an impermeable wall for kinematic reasons, the requirement
of no-slip can be relaxed. In other words, instead of imposing a zero
tangential component of the liquid velocity at the solid, it is
possible to allow for an amount of slippage, described by a slip
length $b$. The slip length for a simple shear flow is the distance
behind the interface at which the liquid velocity extrapolates to zero
\begin{equation}
v_s = b \left[ \partial_z v(z) \right]_{\rm wall},
\label{eq:slip_length}
\end{equation}
where $v_s$ is the tangential velocity at the wall, and the $z$ axis
is perpendicular to the surface. The definition of $b$ is explained in
Fig.~\ref{fig:slip}. It is clear that boundary slip is important
only when the length-scale over which the liquid velocity changes
approaches the slip length. Therefore, it is not surprising that the
slippage effect has not been detected in macroscopic experiments. In
microfluidic devices, however, where the liquid is highly confined,
the boundary slip is important~\cite{vinogradova.oi:1999.a}.

Indeed, water flow in capillaries of small diameter with smooth
hydrophobic walls has been investigated and slip at the wall had to be
postulated to recover the viscosity of
water~\cite{churaev.nv:1984.a,watanabe.k:1999.a}.
These results were confirmed by directly probing the fluid velocity at
a solid surface using total internal reflection-fluorescence
spectroscopy~\cite{pit.r:2000.a,tretheway.dc:2002.a} as well as
 double focus confocal fluorescence
cross-correlation~\cite{lumma.d:2003.a}.
Several indirect methods were also used: the quartz-crystal
microbalance (QCM)~\cite{krim.j:1996.a}, the surface force apparatus
(SFA)~\cite{horn.rg:2000.a,zhu.yx:2001.a,zhu.yx:2002.a,baudry.j:2001.a},
and the atomic force microscope (AFM)~\cite{craig.vsj:2001.a,vinogradova.oi:2003.a}.
The magnitude of the slip length $b$ was sometimes greater than $100 {\rm nm}$
for partially wetted walls~\cite{pit.r:2000.a,zhu.yx:2001.a}. In some
cases the shear rate did not affect the amount of slip in the observed
range~\cite{pit.r:2000.a,vinogradova.oi:2003.a}; in others a strong
dependence on the velocity was
found~\cite{horn.rg:2000.a,zhu.yx:2001.a}. It was also shown that both
surface roughness and strength of the fluid-surface interactions
affect the wall
slip~\cite{churaev.nv:1984.a,zhu.yx:2002.a,vinogradova.oi:2003.a}.

The no-slip condition can also be violated in more complex
systems. Boundary slip has been suggested for polymer
melts~\cite{brochard.f:1992.a,ajdari.a:1994.a,brochardwyart.f:1996.a}
and liquid crystals~\cite{francescangeli.o:1999.a} (for the rotational
motion of molecules).


The origin of such large slippages remains unclear despite
considerable theoretical effort. On the theoretical side, molecular
dynamics simulations have shown that the molecules can slip directly
over the solid due to the fact that the strength of attraction between
the liquid molecules is greater than the competing solid-liquid
interaction~\cite{sun.m:1992.a,sun.m:1992.b,bocquet.l:1993.a,thompson.pa:1997.a,barrat.jl:1999.a,barrat.jl:1999.b}. In general,
wall slip was found on non-wetted surfaces, i.e. when the contact
angle is large. However, the simulation results were not entirely
consistent with the experimental data, by predicting a much lower slip
length~\cite{cieplak.m:2001.a,sokhan.vp:2001.a} and substantial 
slippage only at large contact angles~\cite{barrat.jl:1999.b}. 
It has also been demonstrated that the surface roughness may both
reduce~\cite{cottin_bizonne.c:2003.a} and
increase~\cite{hocking.lm:1976.a,ponomarev.iv:2003.a} the friction of the
fluid past the boundaries.

\begin{figure}
\begin{center}
\includegraphics[width=5cm, angle=0]{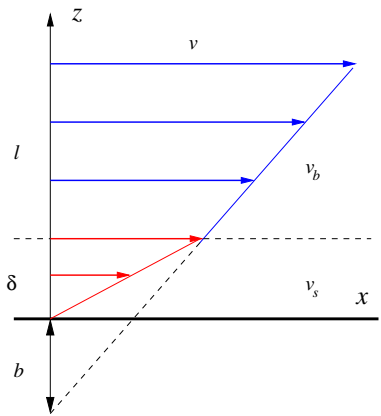}
\end{center}
\caption[slip length]{
\label{fig:geom_bin} 
Slip length $b$ for a binary mixture.
}
\end{figure}

Other ideas invoke the formation of a new phase at the wall. The
possible source of the surface layer could be a gas (lubricant)
dissolved in the liquid, forming bubbles nucleating at the liquid/solid
interface~\cite{ruckenstein.e:1983.a,degennes.pg:2002.a,vinogradova.oi:1995.a}.
Experimental evidence for the formation of a wetting layer under flow
has been found by Tanaka~\cite{tanaka.h:2001.a,tanaka.h.1993.c}.
The boundary layer can have a lower viscosity than the bulk
value~\cite{vinogradova.oi:1995.b}. Tuning the size and the properties
of this layer one can obtain large values of the slip length (see
Fig.~\ref{fig:geom_bin}).

Indeed, in a sharp interface limit, when the width of the interface is
much smaller than the width of the slab and the thickness of the
boundary layer, we can neglect the structure of the interface. Then
the problem is reduced to the shear flow of two phases ($s$ for
``surface'', $b$ for ``bulk'') with viscosities $\eta_s$ and $\eta_b$,
respectively, and thicknesses $\delta$ and $l$, respectively (see
Fig.~\ref{fig:geom_bin}).  Denoting the velocity profiles in the
surface layer and the bulk by $v_s (z)$ and $v_b (z)$, respectively,
and using stick boundary conditions at $z = 0$ and $z = l + \delta$,
we have
\begin{eqnarray}
v_s (0)          & = & 0 ,              \nonumber \\
v_s (\delta)     & = & v_b (\delta)   ,           \\
v_b (l + \delta) & = & v_0 , \nonumber
\end{eqnarray}
where the second equation is the condition for continuity of the flow
field at the interface between the phases, and $v_0$ is the velocity
at the top of the slab.  Furthermore, we use the condition of zero
divergence of the shear stress tensor at the interface,
\begin{equation}
\partial_z \left( \eta(z) \partial_z v(z) \right) = 0,
\label{eq:zerodivergenceshearstress}
\end{equation}
or
\begin{equation}
\eta_s v'_s (\delta) = \eta_b v'_b (\delta) .
\label{eq:velocity}
\end{equation}
The solution in the bulk 
\begin{equation}
v_b(z) = \kappa v_0
\left( \frac{\eta_b}{\eta_s} - 1 + \frac{z}{\delta} \right),
\label{eq:vbulk}
\end{equation} 
with
$\kappa = \left( {\eta_b}/{\eta_s}+{l}/{\delta} \right)^{-1}$,
results in a slip length~\cite{vinogradova.oi:1995.b}
\begin{equation}
b =  \delta \left( \frac{\eta_b}{\eta_s} - 1 \right) .
\label{eq:b_sharp}
\end{equation}
According to Eq.~(\ref{eq:b_sharp}) the boundary slip can be observed
if the viscosity depends on the composition ($\eta_b \ne \eta_s$) and
the less viscous fraction of the liquid wets the walls better than the
more viscous one ($\eta_b > \eta_s$). It is also clear that there are
two ways to obtain a large slip length. First, by having a
macroscopically thick boundary layer, since the slip length has the
same order of magnitude as the thickness of this layer. Second, by
providing large values of the viscosity contrast $\eta_b / \eta_s$,
e.g. for a gas layer~\footnote{when the gas is in the Knudsen regime,
the slip length does not depend on the thickness of the boundary
layer~\cite{degennes.pg:2002.a}.}.


A more realistic description should allow for a prewetting
transition~\cite{binder.k:1983.a,bonn.d:2001.a} for the liquid/gas or
liquid/lubricant mixture and take into account the structure and the
finite width of the interface region.

The aim of this paper is to include these effects and relate the slip
length to the wettability of the walls, composition of the mixture,
and thermodynamic parameters of the system. We show that the
prewetting transition can give rise to a large boundary slip of the
fluid by generating a thick film of a phase-separated `lubricant' at
the wall which has a lower than the bulk fluid viscosity. Indeed, if
we choose typical values, thickness of the wetting layer (thick film)
$\delta \approx 10 \rm nm$, viscosity contrast $\eta_b~:~\eta_s =
3~:~1$, we find $b \approx 20 \rm nm$, i.e. the prewetting film can
indeed give a large slip length. This value can be further
increased. Indeed, if the phase separation occurs upon cooling
(heating), the thickness of the wetting layer increases with the
increase (decrease) of the temperature: for a system with short-range
forces in the vicinity of the wetting temperature $T_w$, the thickness
diverges as~\cite{bonn.d:2001.a} $\delta \propto - \ln \left(
\left\vert T_w - T \right\vert \right)$.

The paper is organized as follows. In Sec.~\ref{sec:model} we outline
the approach that allows us to calculate the phase diagram, order
parameter profiles, and slip length. Section~\ref{sec:results} gives a
summary of the results. Finally, in Sec.~\ref{sec:conclusions}, we
present some brief conclusions.

\section{Theoretical Model}
\label{sec:model}

Phase separation phenomena in binary and polymer mixtures have been
intensively studied by theory, experiment, and simulation. While the
most detailed information is available for the static bulk behavior,
much more interesting phenomena occur when studying the dynamics
\cite{gunton.jd:1983.a,bray.aj:1994.a,wagner.aj:1998.a,kendon.vm:1999.a},
or the behavior near surfaces and in confined geometries
\cite{binder.k:1983.a,dietrich.s:1988.a,binder.k:2003.a,duenweg.b:2003.a}. The
theoretical understanding of phenomena which combine both aspects
(i.~e. dynamics near surfaces) is still at its infancy
\cite{puri.s:1994.a}, while there are many experiments
\cite{tanaka.h:1993.a,tanaka.h.1993.c,vinogradova.oi:1995.b}.

Our theoretical approach splits the problem of shear flow of a binary
mixture near surfaces into several independent tasks. First, we
calculate the {\em equilibrium} order parameter profiles, completely
disregarding the flow. Restriction to equilibrium thermodynamics
allows us to introduce a change of ensemble: we fix the chemical
potential difference (semi-grand-canonical ensemble) instead of fixing
the composition, which is conceptually and computationally easier. The
order parameter profile then results in a viscosity profile, which in
turn allows us to calculate the stationary velocity profile by solving
Eq.~(\ref{eq:zerodivergenceshearstress}), again using stick boundary
conditions at both surfaces. It is not immediately obvious that this
split--up is justified. We discuss the restrictions and underlying
assumptions of this approach in Appendix~\ref{sec:model_h}. Finally,
the slip length is calculated from the stationary velocity profile.

\subsection{Free Energy of a Binary Mixture}
\label{sec:phase}

To describe the bulk phase as well as the interface structure,
`phase-field' models \cite{binder.k:1983.a} are often used. In this
approach the order parameter $\phi$ is introduced. For a 
binary mixture $\phi$ is a composition variable, defined as $\phi =
(n_1 - n_2) / (n_1 + n_2)$, where the $n_i$ are the number densities
of the two species. This order parameter varies slowly in the bulk
regions and rapidly on length scales of the interfacial width. The
unmixing thermodynamics is described via a free energy functional.

Since the material is confined in a container in any experiment, phase
separation is always affected by surface effects
\cite{evans.r:1986.a,evans.r:1987.a,gelb.ld:1997.a,gelb.ld:1997.b,%
gelb.ld:1999.a}. To include them, appropriate surface terms
responsible for the interaction of the liquid with the container walls
are added to the free energy
\cite{cahn.jw:1977.a,binder.k:1983.a,degennes.pg:1985.a,%
dietrich.s:1988.a,bonn.d:2001.a,schick.m:1990.a}.

In the phase-field approach the semi-grand potential of a binary mixture is
written as \cite{bonn.d:2001.a}
\begin{equation}
{\Omega}\{\phi\} = \frac{1}{a^3}
\int d V \left[ \frac{k}{2} a^2 \left(\nabla \phi \right)^2
+ f(\phi) - \mu \phi \right] + \Psi_s ,
\label{eq:grand_pot}
\end{equation}
where $a$ is a normalization length of the order of the size of a
molecule, $f(\phi)$ is the Helmholtz free energy density of the
mixture, while $\mu$ is the chemical potential thermodynamically
conjugate to the order parameter $\phi$, and $\Psi_s$ is the surface
energy.

The explicit form of the Helmholtz free energy $f(\phi)$ varies
depending on the type of mixture. However, the simple observation that the two
phases must coexist implies that there are two minima in the free
energy at the respective values of the order parameter. We here adopt
the mean-field model for a regular (symmetric) mixture
\cite{reichl.le:1998.a,rowlinson.js:1969.a}
\begin{eqnarray}
\nonumber
f(\phi) 
& = & \frac{\chi}{4} (1-\phi^2) + \\
&&    k_B T \left[ 
\frac{1 + \phi}{2} \ln \frac{1 + \phi}{2} + 
\frac{1 - \phi}{2} \ln \frac{1 - \phi}{2}
\right] ,
\label{eq:f_rm}
\end{eqnarray}
where the first term on the right-hand side corresponds to the excess
energy of mixing. Note that this is one of the simplest models to
describe unmixing; a more realistic description would need a more
sophisticated function, which also takes into account a dependence
on the overall density, which can vary along the profile (see
Appendix \ref{sec:model_h}).

The term $\left(\nabla \phi \right)^2$ is needed to provide spatial
structure to the theory: at phase coexistence, there are two bulk
equilibrium order parameter values $\phi_+$ and $\phi_-$ with the same
free energy density, $f(\phi_+) = f(\phi_-)$. Without the gradient
term, a structure with very many interfaces between the $\phi_+$ and
$\phi_-$ phase would be entropically favored. The term $(k/2)
\left(\nabla \phi \right)^2$ is the simplest one which penalizes
interfaces. While this is justified near the critical point, where
interfaces are very wide and the order parameter varies smoothly, a
more realistic description at strong segregation (where the interface
becomes rather sharp) would require higher--order gradients, too.

\subsection{Surface Free Energy}

To describe the interaction with the walls we used the quadratic
approximation for the surface energy
\cite{bonn.d:2001.a,flebbe.t:1996.a}
\begin{equation}
\Psi_s = \frac{1}{a^2} \int \left[ - h \phi_s - 
\frac{1}{2} \gamma \phi_s^2 \right] dS, 
\label{eq:surface_energy}
\end{equation}
where $\phi_s$ is the surface value of the order parameter and the
parameters $h$ and $\gamma$ are referred to as the short-range surface
field and the surface enhancement, respectively. The short-range
surface field, $h$, is a measure of the attractiveness (or
repulsiveness, if negative) of the surface to the component $1$. In
real systems it can be of either sign and of any magnitude. The
surface coupling enhancement, $\gamma$, represents the effect that a
molecule close to the substrate has fewer neighbors than a molecule in
the bulk; $\gamma$ is estimated to be small and negative. We have
implicitly assumed that all surface effects are of short range, thus
$f_s$ depends on the local concentration at the walls only. Because of
long range van der Waals forces this is not fully realistic. However,
for large enough wall separations the differences to the short range
case are rather minor.

We introduce dimensionless units by setting $a = 1$, $\chi = 1$,
and $k_B = 1$. Hence, energies are given in units of $\chi$,
temperatures in units of $\chi / k_B$, and lengths in units of $a$.
Furthermore, a value of $k = 1$ has been used throughout the
calculations of this paper.

\subsection{Euler-Lagrange Equations}

In thermal equilibrium the grand potential (\ref{eq:grand_pot}) must
be minimal. Variation of this functional yields an Euler-Lagrange
equation together with two boundary conditions. Due to translational
symmetry in $x$ and $y$ direction, the problem is one-dimensional.  In
contrast to the situation considered in Introduction, we now focus on
the case of {\em two} identical walls separated by a distance $L$,
which is chosen large enough such that the two wetting layers do not
overlap, and bulk behavior is established in the center of the slab.
It is then convenient to choose the origin of the coordinate system at
the center of the film. In this coordinate system the Euler-Lagrange
equation and boundary conditions read
\begin{eqnarray}
\label{eq:euler}
&&k\frac{\partial^2 \phi}{\partial z^2} + \frac{1}{2} \phi 
- \frac{1}{2} T \ln \frac{1+\phi}{1-\phi} +  \mu = 0, \\
\label{eq:boundary}
&& \pm k\left. \frac{\partial \phi}{\partial z} + h 
+ \gamma \phi \right|_{z = -L/2, L/2} = 0 .
\end{eqnarray}
This boundary-value problem was solved using the relaxation method
\cite{press.wh:1992.a}. In general, it can have up to three different
solutions: one stable, one metastable, and one unstable. The
relaxation method yields only the metastable and the stable
solution. The unstable solution with the highest free energy has
negative response function $(\partial \phi / \partial
\mu)_T$ and is eliminated by the relaxation method automatically.

To select stable solutions, we calculated the grand potential of the
mixture, $\Omega$, for both stable and metastable solutions and chose
the solution with the lowest grand potential. This allows the accurate
determination of the phase diagram.

\subsection{Velocity Profiles and Slip Length}

To calculate the velocity profiles, we assumed that the viscosity of
the system is simply a linear combination of that of the individual
components
\begin{equation}
\eta(z) = \eta_{s} \frac{1 + \phi(z)}{2} + 
          \eta_{b} \frac{1 - \phi(z)}{2},
\end{equation}
where the viscosity contrast between the two components has been
chosen as $\eta_{s}:\eta_{b} = 1:3$. Of course, a more realistic
description would have to introduce a more complicated dependence on
$\phi$, and also take into account the dependence on the overall
density. The stationary velocity profile is the solution of
Eq.~(\ref{eq:velocity})
\begin{equation}
v(z) = v_{\rm w} c^{-1} \int_{-L/2}^{z} \frac{dz}{\eta(z)},
\label{eq:vel_profile}
\end{equation}
where
\begin{equation}
c = \int_{-L/2}^{L/2} \frac{dz}{\eta(z)},
\end{equation}
and we assumed stick boundary conditions at the walls, $v(-L/2) = 0$
and $v(+L/2) = v_{\rm w}$.
The value of the slip length was obtained by fitting the
bulk region of the velocity profile (\ref{eq:vel_profile}) with a
linear regression.

\section{Results and Discussion}
\label{sec:results}

\subsection{Order Parameter Profiles and Prewetting Phase Diagram}

\begin{figure}
\begin{center}
\includegraphics[width=8cm, angle=0]{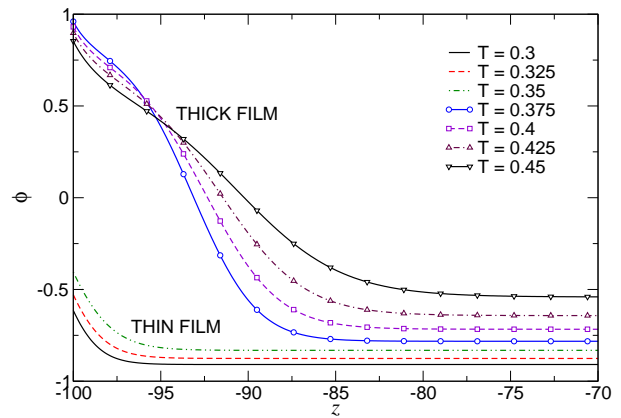}
\end{center}
\caption[Density profiles]{
\label{fig:profile} 
Typical order parameter profiles. Only the part next to the wall is
shown. Thickness of the slab $L=200$, $ \mu = -0.002$, $\gamma =
-0.01$, $h=0.2$. For this chemical potential the thin-thick
film transition occurs at $T \approx 0.35$.
}
\end{figure}

Typical order parameter profiles at the same value of the chemical
potential $\mu = -0.002$ and a set of temperatures are shown in
Fig.~\ref{fig:profile}. Only the profiles corresponding to the stable
thermodynamic state are shown. The profiles have a well-developed flat
region in the center of the film, which indicates that there should
not be finite size effects for the chosen thickness of the slab
($L=200$). This flat region is important for defining the slip length, 
see Eq.~(\ref{eq:slip_length}). This requires a well-defined linear 
velocity profile in the bulk.

In order to understand these profiles, one should note that the very
small value of $\mu$, combined with the low temperature, implies that
the system is rather close to bulk coexistence. Bulk coexistence,
however, is characterized by two equilibrium values of the order
parameter with large absolute value and opposite sign. The small
negative value of $\mu$ singles out the negative order parameter
value, while the absolute value is only slightly changed. Introducing
the surface with preference of the other phase, one obtains a profile
which is slightly bent up near the surface. The prewetting transition
(see Refs.~\cite{cahn.jw:1977.a,schick.m:1990.a}) occurs upon
increasing the temperature, and results in a sudden increase of the
surface excess coverage, i.~e. it is a first-order transition between
a thin-film and a thick-film state. This jump is directly observed at
a temperature near $T=0.35$ in the profiles of Fig.~\ref{fig:profile}.
Taking into account the remarks made in the introduction and
Eq.~(\ref{eq:b_sharp}), we can already anticipate small slip lengths
for thin films and large slip lengths for thick films (above the
prewetting transition temperature). Since the thin-thick film
transition is a first-order transition, one can also expect a jump in
the slip length at the prewetting transition temperature.

\begin{figure}
\begin{center}
\includegraphics[width=8cm]{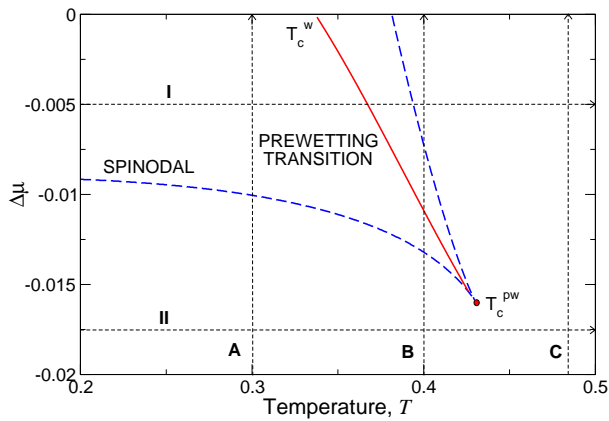}
\end{center}
\caption[Prewetting phase diagram]{
\label{fig:wetting}
Prewetting phase diagram calculated for $h = 0.2$, $\gamma = -0.01$,
$L=200$. The solid line is the prewetting transition line ending at
the prewetting critical point. Metastability limits of thick and thin
films (spinodals) are shown with dashed lines. $\rm A$, $\rm B$, $\rm
C$, $\rm I$, and $\rm II$ are thermodynamic paths used in the text.
}
\end{figure}

To calculate the thin-thick film transition temperature, we followed
the metastable solution up to its stability limit (spinodal)
calculating the grand potential for both stable and metastable
solutions. The transition line was determined from the intersection of
grand potentials of stable and metastable solutions (thick and thin
films). Both spinodals as well as the transition line are shown in a
prewetting transition phase diagram, Fig.~\ref{fig:wetting}. The solid
curve is the prewetting curve ending at the prewetting critical point
and the dashed curves are spinodals or metastability limits of the
metastable states.

\subsection{Slip Length}
\begin{figure}
\begin{center}
\includegraphics[width=8cm]{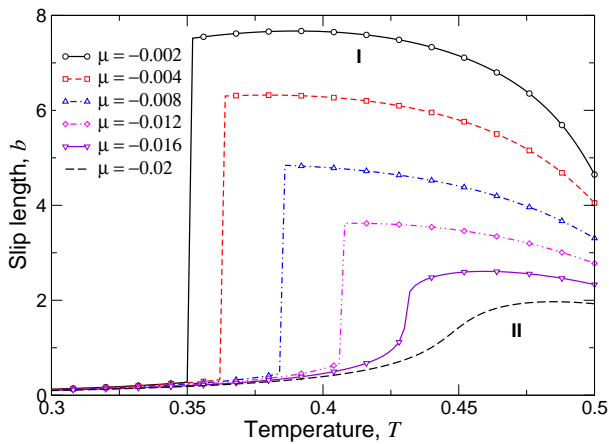}
\end{center}
\caption[Slip length vs temperature]{
\label{fig:slip_T}
Slip length vs. temperature calculated for several values of the
chemical potential.  $h = 0.2$, $\gamma = -0.01$, $L=200$.
}
\end{figure}

\begin{figure}
\begin{center}
\includegraphics[width=8cm]{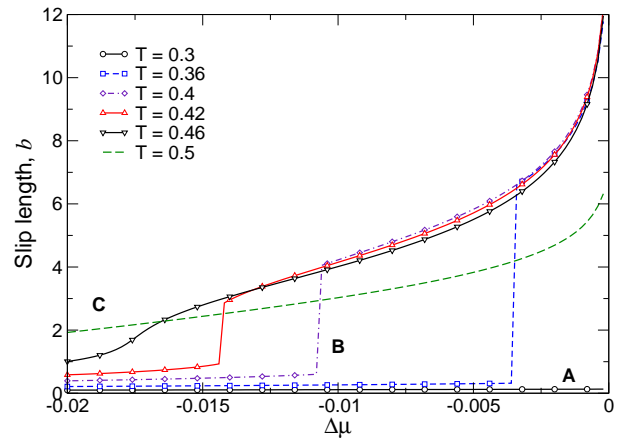}
\end{center}
\caption[Slip length vs chem potential difference]{
\label{fig:slip_mu}
Slip length vs. chemical potential difference. $h = 0.2$, $\gamma =
-0.01$, $L=200$.
}
\end{figure}

\begin{figure}
\begin{center}
\includegraphics[width=8cm]{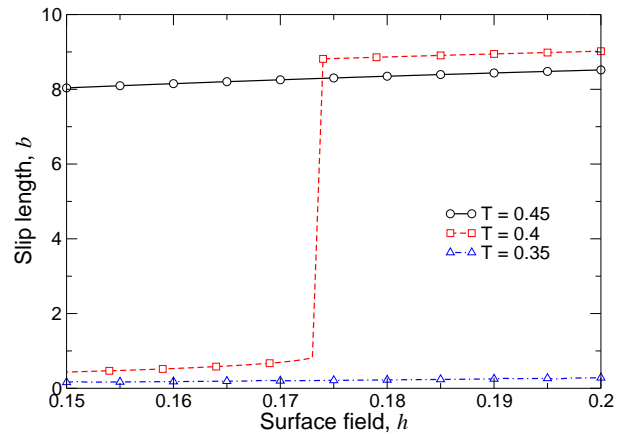}
\end{center}
\caption[]{
\label{fig:slip_h}
Slip length vs. short-range surface field. $ \mu = -0.001$.
}
\end{figure}

The temperature dependence of the slip length for several values of
the chemical potential (or average volume fraction) is presented in
Fig.~\ref{fig:slip_T}. As we anticipated, $b(T)$ has a jump at the
prewetting transition temperature, when the thick film is formed (path
$\rm I$ in Fig.~\ref{fig:wetting}). Before the jump the slip length is
small and practically does not depend on temperature. After the
transition the slip length decreases with the temperature increasing,
even though the prewetting film gets thicker. This is because the bulk
volume fraction (order parameter $\phi$) increases, giving rise to a
decrease in the bulk viscosity (or, equivalently, viscosity contrast)
(see Fig.~\ref{fig:profile}).

If we are above the prewetting critical point (path $\rm II$ in
Fig.~\ref{fig:wetting}) there is no jump-like transition, but rather a
smooth increase of the slip length with the temperature.

The dependence of the slip length on the chemical potential is shown
in Fig.~\ref{fig:slip_mu}. If the temperature is below $T_{\rm c}^{\rm
w}$ then the thin film is always stable and the slip length is small
(path A in Fig.~\ref{fig:wetting}). Once we intersect the prewetting
transition line, the thick film forms with the increase of the average
fraction of the more wettable phase (values of the chemical potential
close to zero) and the slip length jumps abruptly to higher values
(path B in Fig.~\ref{fig:wetting}). Above the prewetting critical
point $T_{\rm c}^{\rm pw}$ the slip length increases monotonically
with the increase of the chemical potential (see path $\rm C$ in
Fig.~\ref{fig:wetting}).

Finally, the dependence of the slip length on the surface field $h$ is
shown in Fig.~\ref{fig:slip_h}. It is clear that there is a threshold
value of the surface field (contact angle) when the thick film is
formed. At this value of the surface field the slip length increases
abruptly. Below the threshold the slip length is small and practically
does not depend on the surface field.

\section{Conclusions}
\label{sec:conclusions}

To summarize, when the prewetting transition occurs in a flow
experiment, it may indeed generate a strong slippage. Prewetting
provides a mechanism of generating a macroscopically thick film at the
wall. If this film has a lower viscosity than the bulk value, a strong
slippage can be observed above the prewetting transition
temperature. The value of the slip length has a jump-like dependence
on temperature, concentration of the phase-separated liquid, and
surface field (contact angle).

The mean-field model of wetting is rather general and can be applied
to liquid-gas systems, binary mixtures, as well as incompressible
polymer mixtures in the long wavelength approximation
\cite{flebbe.t:1996.a}. This implies that the large boundary slip due
to prewetting can be observed in all these systems.

Another goal of this paper is to stimulate accurate quantitative
measurements of the slip length combined with measurements of the
fluid wetting properties. This will allow to study the underlying
microscopic mechanisms of slippage, and choose an adequate model for
every experimental situation. In particular, it would be very
interesting to measure the slip length as a function of temperature
while the system undergoes a prewetting transition, with the adsorbed
species having a lower viscosity. Our results indicate that in such a
case the slip length should undergo an abrupt change, and we believe
that this will be true beyond the limitations of our idealized model.

\appendix

\section{Hydrodynamics of Slippage in Binary Fluids}

\label{sec:model_h}

We consider a binary fluid (species $1$ and $2$) confined between two
infinite parallel planes to a thin slab of thickness $L$. The plane
normal is in the $z$ direction. We denote the mass densities of the
two species by $\rho_1$ and $\rho_2$, respectively, and introduce the
linear combinations $\rho = \rho_1 + \rho_2$ (total mass density) and
$\Delta \rho = \rho_1 - \rho_2$.

Concerning the thermodynamics of the system, we notice that the
pressure tensor $p_{\alpha \beta}$ (Greek letters denote Cartesian
indices) is anisotropic as a result of the finite size effect 
and the interface contribution to the free energy 
\cite{allen.mp:2000.a}. Since the system is fluid, there is no 
elastic response to shear, and hence $p_{\alpha \beta} = 0$ for
$\alpha \ne \beta$. Furthermore, $p_{xx} = p_{yy}$ for symmetry
reasons. Apart from this anisotropy, we assume that there is no
further inherent anisotropy. In particular, we assume that all
transport coefficients (interdiffusion coefficient, thermodiffusion
coefficient, viscosities, etc.) retain their simple scalar nature as
in the isotropic macroscopic bulk fluid. The viscous stress tensor is
hence written as
\begin{equation}
\sigma_{\alpha \beta} 
= \eta \left( \partial_\alpha v_\beta + \partial_\beta v_\alpha 
  - \frac{2}{3} \delta_{\alpha \beta} \partial_\gamma v_\gamma \right)
+ \zeta \delta_{\alpha \beta} \partial_\gamma v_\gamma ,
\end{equation}
where the Einstein summation convention is implied, $\bm v$ denotes
the fluid flow velocity, while $\eta$ and $\zeta$ are the shear and
the bulk viscosity, respectively. We assume that these parameters
depend on density and composition, i.~e. $\eta = \eta( \rho, \Delta
\rho)$ and $\zeta = \zeta (\rho, \Delta \rho)$.

We consider the situation that the fluid is weakly driven such that a
shear flow in $x$ direction develops, with shear gradient in $z$
direction. In this limit of weak driving, it is reasonable to assume
that no symmetry is broken except translational invariance in $z$
direction. The system remains translationally invariant in $x$ and $y$
direction, and gradients (of any quantity) occur only in $z$
direction. We also assume that the system is kept at constant
temperature throughout, i.~e. that the heat production is negligibly
small. This is reasonable for small shear rates (note that the heat 
production is proportional to the square of the shear rate). In addition, 
the heat conductivity is quite large for many real fluids.

Under these assumptions, we seek a stationary solution of the
hydrodynamic equations of motion \cite{landau.ld:1995.a} for the
outlined geometry and symmetry. Since the velocity flow field $\bm v
= (v_x, 0, 0)$ is defined as the center--of--mass velocity of a volume
element, the dynamics of $\rho$ is governed by pure convection:
\begin{equation} \label{eq:continuity}
\partial_t \rho + \partial_\alpha \left( \rho v_\alpha \right) = 0.
\end{equation}
A stationary solution implies $\partial_t \rho = 0$, while our
geometry leads to $\partial_\alpha \left( \rho v_\alpha \right) =
\partial_z \left( \rho (z) v_z \right) = 0$. The continuity equation
is therefore identically fulfilled, and a non--constant mass density
profile $\rho(z)$ is permitted.

For the density difference there is also interdiffusion:
\begin{equation}
\partial_t \Delta \rho + 
\partial_\alpha \left( \Delta \rho v_\alpha \right) = 
- \partial_\alpha j_\alpha ,
\end{equation}
where $\bm j$ is the interdiffusion current. Again, the left hand
side vanishes identically for our flow. Furthermore, the interdiffusion
current vanishes in the stationary state:
\begin{equation}
\bm j = 0.
\end{equation}
The (full nonlinear) Navier--Stokes equation is written as
\begin{equation}
\partial_t \left( \rho v_\alpha \right) +
\partial_\beta \left( \rho v_\alpha v_\beta \right) =
- \partial_\beta p_{\alpha \beta} + 
\partial_\beta \sigma_{\alpha \beta} .
\end{equation}
For our flow, the left hand side vanishes identically, and hence
\begin{eqnarray} \label{eq:psigma}
    \partial_z p_{zz}      & = & 
    \partial_z \sigma_{zz}   = 0 \label{eq:constpress}, \\
0 = \partial_z p_{xz} (z)  & = & 
    \partial_z \sigma_{xz}   =  
    \partial_z \left( \eta \partial_z v_x  \right), 
                                 \label{eq:velocityprofile}   \\
0 = \partial_z p_{yz}      & = & 
    \partial_z \sigma_{yz} = 0.
\end{eqnarray}
From Eq.~(\ref{eq:constpress}) one sees that the pressure profile
$p_{zz}$ must be constant, while Eq.~(\ref{eq:velocityprofile})
allows to obtain the velocity profile via integration, as soon
as the viscosity profile $\eta(z)$ is known from the profiles
$\rho(z)$ and $\Delta \rho (z)$.

We now turn to the constitutive equation for the interdiffusion
current $\bm j$. In non--equilibrium thermodynamics, the dissipative
currents are assumed to be linear in the gradients of the intensive
thermodynamic variables. A binary system has three independent
thermodynamic variables, for which we can take any appropriate
set. For our purposes it is particularly useful to choose the pressure
$p$, the temperature $T$, and the chemical potential $\mu$, which is
defined as the variable which is thermodynamically conjugate to the
order parameter $\phi = (n_1 - n_2) / (n_1 + n_2)$, where the $n_i$
denote the particle number densities (see main text). Macroscopically,
the constitutive equation would thus read
\begin{equation}
j_\alpha = - \Gamma_1 \partial_\alpha \mu
           - \Gamma_2 \partial_\alpha T
           - \Gamma_3 \partial_\alpha p ,
\end{equation}
where $\Gamma_i$ are suitable scalar Onsager coefficients. Note that
the pressure gradient may appear since $\mu$ is not the variable
conjugate to $\Delta \rho / \rho$ (which is often used in the
literature \cite{landau.ld:1995.a}), but rather to $\phi$. We now
generalize this equation to the case of an anisotropic pressure
tensor, but retain the scalar nature of the Onsager coefficients
in accordance with our assumptions (similar to what is done in
model H \cite{kendon.vm:2001.a,zhang.zh:2001.a}). We thus find
\begin{equation}
j_\alpha = - \Gamma_1 \partial_\alpha \mu
           - \Gamma_2 \partial_\alpha T
           - \Gamma_3 \partial_\beta p_{\alpha \beta} ,
\end{equation}
or, taking into account that there are only gradients in $z$ direction,
and that $\bm j$ vanishes,
\begin{equation}
0 = - \Gamma_1 \partial_z \mu
    - \Gamma_2 \partial_z T
    - \Gamma_3 \partial_z p_{zz} .
\end{equation}
Now, $\partial_z T$ vanishes due to our assumption of an isothermal
system, while $\partial_z p_{zz}$ vanishes as a consequence of the
Navier--Stokes equation (see Eq.~\ref{eq:constpress}). For this
reason, the profile of the chemical potential, $\mu (z)$, must
be a constant, too.

In summary, we find that under the given assumptions (isothermal
system, isotropic Onsager coefficients, weak driving) the conditions
for the stationary state are identical to those in thermal equilibrium
(all intensive variables must have constant profiles). This permits to
first calculate the density profiles just as equilibrium profiles,
completely disregarding the flow, and then, in a second step, to
calculate the velocity profile by solving
Eq.~(\ref{eq:velocityprofile}). This has been done in the main text
for a simple model for the unmixing thermodynamics, and a linear
dependence of the viscosity $\eta$ on the order parameter. Note also
that the restriction to pure equilibrium thermodynamics allows us to
introduce a change of ensemble: instead of considering the composition
as fixed, we rather view $\mu$ as fixed (semi--grand ensemble), which
is conceptually and computationally easier. Of course, a quantitative
comparison with experiments is not possible for such simple models;
one would have to use much more sophisticated free energy functionals,
and a better model for the concentration dependence of $\eta$ as well.
Furthermore, one should expect that at moderate shear rates only
 hydrodynamic instabilities (e.~g. bubble formation near the
surfaces) should occur which invalidate the assumption of
translational invariance in $x$ and $y$ direction.

\acknowledgments

We are grateful to R.~Evans, F.~Feuillebois, K.~Kremer, M.~M\"uller,
and J.~Vollmer for useful discussions. DA acknowledges the support of
the Alexander von Humboldt Foundation.


\end{document}